\documentclass[english,aps,prb,superscriptaddress]{revtex4}
\pdfoutput=1
\usepackage[T1]{fontenc}
\usepackage[latin9]{inputenc}
\usepackage{textcomp}
\usepackage{graphicx}
\usepackage{amssymb}

\makeatletter

\providecommand{\boldsymbol}[1]{\mbox{\boldmath $#1$}}

\usepackage{mdwlist}

\usepackage{babel}
\makeatother

\begin{document}

\title{The Glass Transition and the Jarzynski Equality}

\author{Stephen R. Williams}

\affiliation{Research School of Chemistry, The Australian National University,
Canberra, ACT 0200, Australia}

\author{Debra J. Searles}

\affiliation{Nanoscale Science and Technology Centre, School of Biomolecular and
Physical Sciences, Griffith University, Brisbane, Qld 4111, Australia}

\author{Denis J. Evans}

\affiliation{Research School of Chemistry, The Australian National University,
Canberra, ACT 0200, Australia}

\begin{abstract}
A simple model featuring a double well potential is used to represent
a liquid that is quenched from an ergodic state into a history dependent
glassy state. Issues surrounding the application of the Jarzynski
Equality to glass formation are investigated. We demonstrate that
the Jarzynski Equality gives the free energy difference between the
initial state and the state we would obtain \emph{if} the glass relaxed
to true thermodynamic equilibrium. We derive new variations of the
Jarzynski Equality which are relevant to the history dependent glassy
state rather than the underlying equilibrium state. It is shown how
to compute the free energy differences for the nonequilibrium history
dependent glassy state such that it remains consistent with the standard
expression for the entropy and with the second law inequality. 
\end{abstract}

\date{\today}

\maketitle

\section{Introduction}

In many real chemical and physical systems the observed distribution
of the components is not that expected by statistical mechanics for
a system at thermodynamic equilibrium. This can occur in the absence
of mechanical forces driving the system away from equilibrium. Many
compounds exist as more than one polymorph at standard room temperature
and pressure, despite one polymorph having a significantly lower free
energy than the others. Fluids can be cooled below the temperature
at which thermodynamics predicts that a solid phase would be thermodynamically
more stable \citep{Debenedetti-book}. Some components exist in different
abundance than that predicted thermodynamically. These situations
sometimes persist for timescales that are longer than human measurement
allows or even for geological timescales. This behaviour might simply
occur due to a slow transformation to the more stable state, or extreme
rarity of necessary nucleating events might mean that the system is
trapped in some nonergodic state for timescales that are incredibly
long. Systems in these states are often considered as being in `metastable'
states, although it is perhaps inappropriate terminology for a polymorph
like diamond (an allotrope of carbon that has higher free energy than
graphite) which would normally be considered quite stable. Such systems
are also often described as being in `nonequilibrium states', but
they are nondissipative and no mechanical force is applied to prevent
relaxation to the equilibrium distribution. To simplify terminology
we will refer to these \emph{history dependent, non-dissipative, nonergodic,
time independent nonequilibrium states} as \emph{quasiequilibrium}
states.

In the past it has often been assumed that subsets of the components
will be equilibrated, and their relative distributions will be given
by equilibrium, Boltzmann distributions. Such ideas have been exploited
in the so called energy landscape picture of the glass transition
\citep{Stillinger-PRA-82,Stillinger-sci-1984,Debenedetti-Nat-01,Sciortino-JSM-05}.
It has also been assumed that equilibrium thermodynamics can be applied
to these systems, which requires that the phase space domains of the
subsets do not change with small changes in the state point. 

Quasiequilibrium states can be formed in various ways - \emph{e.g.}
by temperature quenching, changing the potential energy function or
changing the pressure of a system so rapidly that the system is not
given sufficient time to adjust to the new conditions and the inter-domain
weights can therefore not be expected to be Boltzmann. If the barrier
for transformation between two or more local minima is high, the non-equilibrium
distribution between the phase space domains will persist. The relative
distribution between the domains will depend on the way they are prepared
and will therefore not be a Boltzmann distribution. Nevertheless,
once trapped, there is ample time for the subsystems to become equilibrated
within their restricted phase space subdomains. Williams and Evans
\citep{Williams-JCP-07} produced convincing arguments that within
these individual ergodic phase space sub-domains the internal distribution
of states is given by a Maxwell-Boltzmann distribution and, using
the fluctuation theorem as a sensitive test of aspects of the domain
statistics, they confirmed that in aged glasses that are not too close
to the glass transition, the intra-domain statistics are Boltzmann
and the domains are robust with respect to small but finite changes
in the external thermodynamic state variables (temperature, pressure
etc.) \citep{Evans-PRL-93-ECM2,Evans-PRE-94,Evans-Adv.-Phys.-02}. 

A particularly interesting example of a quasiequilibrium system is
a glass. When a good glass former is prepared, it is not able to relax
to true thermodynamic equilibrium for an extraordinarily long time,
often many thousands of years and in the case of the natural glass
obsidian, some hundred million years. The system remains in a very
long lived, history dependent, quasiequilibrium state. Nevertheless,
from a macroscopic point of view, the material appears to be an ordinary
equilibrium solid. The fundamental thermodynamics and statistical
mechanics of glass is a topic of active research. 

Here we consider a simple model that could be used to represent a
glass. Unlike some solids (e.g. allotropes of carbon), glasses have
a structure which resembles that of a liquid. Because of the numerous
long lived structures it is necessary to examine the distribution
of states within the glassy system. However, the model presented below
might also be considered to represent many other systems e.g. a protein
might be frozen into a particular conformation, and we might be interested
in the free energy of this conformation compared to the overall free
energy. 

In recent years the equality for determination of free energy differences
introduced by Jarzynski \citep{Jarzynski-PRL-1997,Jarzynski-PRE-1997}
has received considerable attention. This remarkable equality allows
the difference in free energy between two states in thermodynamic
equilibrium to be computed from an ensemble of nonequilibrium trajectories
or pathways, of finite duration, which transform between the two equilibrium
states. It is an interesting question as to whether this equality
can shed light on a system which is quenched into a glassy state.
It would also be interesting if this equality could be used to compute
the difference in free energy of two polymorphs of the same compound,
or to find the coexistence point in a phase transition. We investigate
how the Jarzynski Equality can be extended to treat these systems.
Since the Jarzynski Equality relates the free energies of different
canonical states at the same temperature, we consider the formation
of a glass by changing the potential energy surface of the system
while keeping temperature constant. This models the formation of glassy
systems by, say, altering the molecular interactions by changing the
pH, or increasing the mole fraction of free polymer in a dispersion
to form a glassy colloidal system \citep{Pham-S-2002}.

\section{Theory}

\subsection{Jarzynski Equality}

Here we will outline a very general approach that can be applied to
arbitrary ensembles and dynamics (deterministic or stochastic) \citep{Reid-EPL-05,Williams-PRL-08}.
It can be used to obtain the Jarzynski Equality (JE) under particular
conditions, but is more general and will be useful in the study of
the quasiequilibrium states in the next section. Consider two closed
N-particle systems: (1) and (2) with arbitrary \emph{equilibrium}
distribution functions. A protocol and the corresponding time-dependent
equations of motion are defined to transform system (1) to system
(2). The dynamics may be thermostatted as in Eq. (\ref{eq:EOM}) below
or it may be isoenergetic or even unthermostatted. We define a generalised
dimensionless {}``work'' $\Delta X_{\tau}(\boldsymbol{\Gamma};0,\tau)$,
that is evaluated for a trajectory of duration ${\normalcolor {\normalcolor \tau}}$, 

\begin{equation}
\exp[\Delta X_{\tau}(\mathbf{\Gamma})]\equiv\frac{P_{eq}^{(1)}(d\mathbf{\Gamma})\: Z^{(1)}}{P_{eq}^{(2)}(d\mathbf{\Gamma}(\tau))\: Z^{(2)}}=\frac{f_{eq}^{(1)}(\mathbf{\Gamma})d{\bf \Gamma}\: Z^{(1)}}{f_{eq}^{(2)}(\mathbf{\Gamma}(\tau))d{\bf \Gamma}(\tau)\: Z^{(2)}}\label{generalised work_distfns}\end{equation}
where $Z^{(i)}$ is the partition function for equilibrium system
$i$ and $P_{eq}^{(i)}(d\mathbf{\Gamma})=f_{eq}^{(i)}({\bf \Gamma})d{\bf \Gamma}$
is the probability of observing the infinitesimal phase volume $d\mathbf{\Gamma}$,
centred on the phase vector $\mathbf{\Gamma}$, according to the \emph{$i^{th}$
}equilibrium distribution function, $f_{eq}^{(i)}$. The phase volume
$d{\bf \Gamma}(\tau)$ is generated from $d{\bf \Gamma}$ using the
equations of motion that take the system from equilibrium state (1)
\emph{towards} state (2) (using the forward protocol). For $\Delta X_{\tau}(\mathbf{\Gamma})$
to be well defined requires that

\begin{enumerate}
\item if $f_{eq}^{(1)}({\bf \Gamma})\ne0$ then $f_{eq}^{(2)}({\bf \Gamma}(\tau))\ne0$,
and \label{enu:ergoconsistGenWork1}
\item the converse, namely that if $f_{eq}^{(2)}({\bf \Gamma}(\tau))\ne0$,
then $f_{eq}^{(1)}({\bf \Gamma})\ne0$. \label{enu:ergoconsistGenWork2}

\suspend{enumerate}The second condition is also required because
if the numerator of Eq. (\ref{generalised work_distfns}) is zero,
the {}``work'' which is the logarithm of the right hand side of
Eq. (\ref{generalised work_distfns}) will not be defined. We call
these conditions the \emph{ergodic consistency conditions for the
generalised work}. They are analogous to the ergodic consistency condition
for the Evans-Searles Fluctuation Theorem\citep{Evans-Adv.-Phys.-02}.

It is trivial to prove that the exponential average of $-\Delta X_{\tau}$
satisfies the following relation:

\begin{equation}
\left\langle \exp(-\Delta X_{\tau})\right\rangle _{eq1}=\int_{\mathbf{\Gamma}|f_{eq}^{(1)}({\bf \Gamma})\ne0}d\mathbf{\Gamma}f_{eq}^{(1)}(\mathbf{\Gamma})\frac{f_{eq}^{(2)}(\mathbf{\Gamma}(\tau))\left\Vert \partial\mathbf{\Gamma}(\tau)/\partial\mathbf{\Gamma}\right\Vert Z^{(2)}}{f_{eq}^{(1)}(\mathbf{\Gamma})Z^{(1)}}=\frac{Z^{(2)}}{Z^{(1)}}\label{generalised Jarzynski}\end{equation}
where the brackets $\left\langle \ldots\right\rangle _{eq1}$ denote
an ensemble average over the initial (i.e. $f_{eq}^{(1)}$) equilibrium
distribution and $d{\bf \Gamma}(\tau)/d{\bf \Gamma}=\left\Vert \partial\mathbf{\Gamma}(\tau)/\partial\mathbf{\Gamma}\right\Vert $.
This relationship is very general \citep{Reid-EPL-05} and shows how
free energy differences can be computed from path integral information
taken from nonequilibrium paths. These paths do not need to be quasistatic.
We call this equality Eq. (\ref{generalised Jarzynski}) the Generalised
Jarzynski Equality (GJE).

The restriction of the integral to those regions where $f_{eq}^{(1)}({\bf \Gamma})\ne0$
means that one completely avoids divergences in the function being
averaged. The validity of Eq. (\ref{generalised Jarzynski}) only
requires:\resume{enumerate}

\item an integrable region in the phase space of the final equilibrium distribution
for which $f_{eq}^{(2)}({\bf \Gamma}(\tau))\ne0$. \label{enu:ErgoConsistGJE}

\suspend{enumerate}We call this the ergodic consistency condition
for the GJE. This condition is more general than the corresponding
ergodic consistency condition for the generalised work. The following
example illustrates a case where this condition breaks down and where
the GJE fails. Consider the adiabatic transformation of one Hamiltonian
system into a different Hamiltonian for which $H(\mathbf{\Gamma}(\tau))\neq H(\mathbf{\Gamma}(0))\;\forall\mathbf{\Gamma}$.
If the two equilibrium states are microcanonical and they have the
same energy then we will have $\forall\mathbf{\Gamma},\; f_{eq}^{(2)}({\bf \Gamma}(\tau))=0$.
So in this example ergodic consistency is violated for both the generalised
work and for the GJE.

Of course we also make the usual physical assumptions that the dynamics
is such that there are no singularities in the equations of motion
so that the trajectories in phase space are well defined and that
the Jacobian in Eq. (\ref{generalised Jarzynski}) is non divergent. 

To obtain the Jarzynski Equality we consider the special case of transformations
using thermostatted dynamics between canonical equilibrium states
with the same temperature. In order to determine the free energy difference,
we consider an ensemble of initial equilibrium states at time $t=0$
that is transformed to a new state over a period $0<t\le\tau$. During
this period, the ensemble of states is not at equilibrium, but if
the transformation is halted at $t=\tau$, the system will eventually
relax to a new equilibrium state. The simplest case involves a change
in the functional form of the internal energy of the system during
the period $0<t\le\tau$ from $H_{0}^{(1)}({\bf \Gamma)}$ to $H_{0}^{(2)}({\bf \Gamma)}$,
after which it is fixed at $H_{0}^{(2)}({\bf \Gamma)}$. We imagine
that while these changes occur to the system of interest that it may
be in contact with a very large heat reservoir, ensuring that the
two equilibrium states are at the same temperature. If we make the
system of interest remote from this reservoir, then it cannot possibly
know the details of how the reservoir operates. As an example, we
can model the remote reservoir by a Gaussian isokinetic reservoir
where the kinetic energy of the reservoir particles is fixed at the
value, $K_{therm}=3N_{therm}k_{B}T/2$ where $T$ is the equilibrium
thermodynamic temperature of the reservoir, $N_{therm}$ is the number
of thermostatted reservoir particles\citep{evans-and-morriss-book,Williams-PRE-04,Williams-MP-2007}.
It is assumed that the thermostat temperature $T$ is identical to
the temperature of the two canonical ensembles between which we wish
to calculate free energy differences. This reservoir can be regarded
as being in thermodynamic equilibrium because it is assumed to have
many more degrees of freedom than the system of interest. 

The equations of motion for the system during the time $0<t\le\tau$
are written as\begin{eqnarray}
\dot{\mathbf{q}}_{i} & = & \frac{\mathbf{p}_{i}}{m}\nonumber \\
\dot{\mathbf{p}}_{i} & = & -\frac{\partial H_{0}(\mathbf{q},\mathbf{p},\lambda(t))}{\partial\mathbf{q}_{i}}-\alpha S_{i}\mathbf{p}_{i}\nonumber \\
\alpha & = & \frac{\Sigma_{i=1}^{N}S_{i}\mathbf{F}_{i}\cdot\mathbf{p}_{i}}{\Sigma_{j=1}^{N}S_{j}\mathbf{p}_{j}\cdot\mathbf{p}_{j}}\label{eq:EOM}\end{eqnarray}
where ${\bf F}_{i}=-\frac{\partial H_{0}(\mathbf{q},\mathbf{p},\lambda(t))}{\partial\mathbf{q}_{i}}$.
In these equations $\lambda$ is a parametric function such that $H_{0}(\lambda(0))=H_{0}^{(1)}$
and $H_{0}(\lambda(\tau))=H_{0}^{(2)}$ and the function $\lambda(t)$
defines the transformation protocol. The switch, $S_{i}$, is defined
such that $S_{i}=1$ for particles that form the thermostatting reservoir
and $S_{i}=0$ when they are part of the system of interest. The variable
$\alpha$ is a Gaussian thermostat multiplier \citep{evans-and-morriss-book}
that fixes the kinetic energy of the reservoir particles. It is easy
to see that for such a system $\dot{H}_{0}^{therm}(\boldsymbol{\Gamma},t)=-2K_{therm}\alpha(\boldsymbol{\Gamma},t)=\dot{Q}$
where $K_{therm}=\Sigma_{i=1}^{N}S_{i}\,\mathbf{p_{i}^{2}}/2m$ is
the kinetic energy of the reservoir particles and $\dot{Q}$ is the
rate at which heat is exchanged with the synthetic thermostat.

In this case the Liouville equation states: $\frac{df}{dt}=-\Lambda f=3N_{therm}\alpha f$,
where $\Lambda=\frac{\partial}{\partial{\bf \Gamma}}\cdot\dot{{\bf \Gamma}}$
is the phase space compression factor \citep{evans-and-morriss-book,sevick-ARPC-2007}.
Hence \begin{eqnarray}
\left\Vert \frac{\partial\mathbf{\Gamma}(\tau)}{\partial\mathbf{\Gamma}}\right\Vert  & = & \frac{f_{eq}^{(1)}({\bf \Gamma})}{f_{\tau}^{(1)}({\bf \Gamma}(\tau))}=\exp\left[\int_{0}^{\tau}dt\,\Lambda(\mathbf{\Gamma}(t))\right]\nonumber \\
 & = & \exp\left[-3N_{therm}\int_{0}^{\tau}dt\,\alpha({\bf \Gamma}(t))\right]=\exp\left[\beta\int_{0}^{\tau}dt\;\dot{H}_{0}^{therm}(\mathbf{\Gamma}(t))\right],\label{eq:Jacobian-1}\end{eqnarray}
where $f_{\tau}^{(1)}$ denotes $f_{eq}^{(1)}$ evolved for a period
$\tau$. In general $f_{\tau}^{(1)}$ is not an equilibrium distribution. 

If the equilibrium distributions $f_{eq}^{(1)}({\bf \Gamma})$ and
$f_{eq}^{(2)}({\bf \Gamma})$ are canonical and at the same temperature,
it is trivial to show using Eq. (\ref{generalised work_distfns}),
that $\Delta X_{\tau}/\beta$ is the total energy change in the system
minus the energy (i.e. the heat) gained by the system from the thermostat
(usually a negative quantity), $-\Delta Q(\boldsymbol{\Gamma};0,\tau)=-\int_{0}^{\tau}dt\;\dot{H}_{0}^{therm}({\bf \Gamma}(t))]$.
That is, using Eqs. (\ref{generalised work_distfns}) \& (\ref{eq:Jacobian-1}),
we see that \begin{eqnarray*}
\Delta X_{\tau}(\boldsymbol{\Gamma};0,\tau) & =\beta & \int_{0}^{\tau}dt\,[\dot{H}_{0}^{tot}(\boldsymbol{\Gamma},t)-\dot{H}_{0}^{therm}(\boldsymbol{\Gamma},t)]=\beta\int_{0}^{\tau}dt\,[\dot{H}_{0}^{tot}(\boldsymbol{\Gamma},t)]-\beta\Delta Q_{\tau}(\boldsymbol{\Gamma};0,\tau)\end{eqnarray*}
\begin{equation}
=\beta\int_{0}^{\tau}dt\,\dot{H}_{0}^{ad}(\boldsymbol{\Gamma},t)=\beta\Delta W_{\tau}(\boldsymbol{\Gamma};0,\tau).\label{work}\end{equation}
Here $\dot{H}_{0}^{ad}$ is the adiabatic (unthermostatted) time-derivative
of the internal energy \citep{endnote-1}. The final equality is obtained
by consideration of the First Law of Thermodynamics, and shows that
in this case (thermostatted dynamics with canonical initial and final
distributions), $\Delta X_{\tau}$ is just the work performed on the
system in the transformation multiplied by $\beta$: $\Delta X_{\tau}(\boldsymbol{\Gamma};0,\tau)=\beta\Delta W_{\tau}(\boldsymbol{\Gamma};0,\tau)$
\citep{Jarzynski-PRE-1997,Evans-MP-2003,Williams-CRP-2007,sevick-ARPC-2007,Reid-EPL-05}.
Substitution of Eq. (\ref{work}) into Eq. (\ref{generalised Jarzynski})
then gives the well known Jarzynski Equality, 

\begin{equation}
\left\langle \exp(-\beta\Delta W_{\tau})\right\rangle _{eq1}=\frac{Z^{(2)}}{Z^{(1)}}=exp[-\beta\Delta A],\label{JE}\end{equation}
where the partition functions $Z^{(i)}$ are related to the Helmholtz
free energy by the equation\begin{equation}
A=-k_{B}T\ln\left(\int d\boldsymbol{\Gamma}\,\exp\left(-\beta H_{0}(\mathbf{\Gamma})\right)\right)=-k_{B}T\ln Z.\label{Helmholtz}\end{equation}

Eq. (\ref{JE}) provides a way of determining the difference in the
Helmholtz free energy, $\Delta A=A^{(2)}-A^{(1)}$, between two canonical
equilibrium states with partition functions $Z^{(1)}$ and $Z^{(2)}$
by measuring the work, $\Delta W_{\tau}$ done over a period $\tau$,
for an ensemble of nonequilibrium pathways starting in state {}``1''
and moving towards but not actually reaching equilibrium state {}``2''. 

The same result is obtained if the initial ensemble is canonical and
the dynamics is either thermostatted by a Nos\'e-Hoover thermostat
or the dynamics are adiabatic. For other ensembles Eq. (\ref{generalised work_distfns})
may not refer to work (see \citep{Reid-EPL-05}). For example the
microcanonical ensemble with the same energy $H_{0}$ at times $t=0$
and $t=\tau$, the generalised {}``work'' $\Delta X_{\tau}$, is
in fact the change in heat.

In the derivation of Eq. (\ref{JE}) it is assumed that the initial
distribution is given by the full canonical ensemble. The initial
distribution must be a fully relaxed ergodic equilibrium state. The
identity is then a mathematical relation about how the free energy
difference on the right hand side is related to various integrals
on the left hand side. If at the end of the protocol for changing
the Hamiltonian the system is not in true thermodynamic equilibrium,
as long as the ergodic consistency conditions hold, subsequent relaxation
to equilibrium does not matter. If ergodic consistency fails (because
the observed phase density at time $\tau$, has \emph{no} overlap
with the final equilibrium distribution) then the Jarzynski Equality
fails. Also in any practical implementation of any GJE the generalised
work needs to be properly defined, so the ergodic consistency condition
for the generalised work takes precedence over that for the GJE itself.

A necessary condition for Eq. (\ref{JE}) or Eq. (\ref{generalised Jarzynski})
to yield correct results in practise is that in the ensemble averaging
process the time reversed path of the most probable path, must be
observed. If the averaging process is not sufficiently exhaustive
for the initial points of these possibly extremely rare events to
be sampled from the initial equilibrium distribution, numerical evaluation
of Eq. (\ref{JE}) or Eq. (\ref{generalised Jarzynski}) will give
misleading results. One can easily see that this is the case. Write
the work (or the generalised work) as the sum of the reversible work
and the purely irreversible work. As the reversible work is just the
free energy difference it can be taken through the average of the
negative exponential. The average of the negative exponential of the
purely irreversible work must now average to unity. This is just the
Nonequilibrium Partition Identity \citep{evans-and-morriss-book,Carberry-JCP-2004}.
It is well known from the Evans Searles Fluctuation Theorem\citet{Evans-Adv.-Phys.-02}
that the necessary condition for this to hold in sampled data is to
see the anti trajectories of the most probable trajectories for the
process considered. This observation has an immediate impact on the
calculation of free energy differences in the thermodynamic limit.
These differences must be calculated for finite systems for a series
of system sizes and then extrapolation must be employed in order to
take the thermodynamic limit.

\subsection{Quasiequilibrium free energies from the quasiequilibrium partition
function}

Consider an ensemble of glass-forming systems at equilibrium at $t=0$.
We then quench the system to a quasiequilibrium glass state by changing
$H_{0}$ over a period $0<t\le\tau$. After the relaxation of transients,
at $t=\tau_{qe}\ge\tau$, we assume that the ensemble remains in the
glass state for a prolonged period of time during which the average
properties of the system seem constant on the time scale of observation.
The effect of this process on the phase space distribution is shown
schematically in Fig. \ref{fig:schematic}. If the time scale over
which the system relaxes to equilibrium is very slow ($t\gg\tau_{qe}$)
we may accurately model the ensemble's distribution function at $t=\tau_{qe}$
by treating it as consisting of a set of non-overlapping phase space
domains $\{D_{\alpha};\alpha=1,N_{D}\}$. These domains partition
the phase space available to any individual sample. By definition
any sample belongs to one and only one phase space domain. Within
individual domains the samples are ergodic (by definition) and time
averaged properties are equal to ensemble averages over sets of samples
belonging to the same phase space domain. The domains have zero overlap
- otherwise they would not be ergodic. When viewed separately each
of these domains appears to be in equilibrium with internal weights
given by the relative Boltzmann weights. However the relative number
of ensemble members in each of these domains is not consistent with
an equilibrium Boltzmann distribution \citep{Williams-JCP-07}. Instead
these relative populations are influenced by the details of the quench
and subsequent ageing process that was used to prepare the ensemble
of samples - they are history dependent. We call this ensemble a quasiequilibrium
ensemble. The phase space distribution function for this ensemble
has been derived by Williams and Evans \citep{Williams-JCP-07}.

\begin{figure}
\includegraphics[clip,scale=0.5]{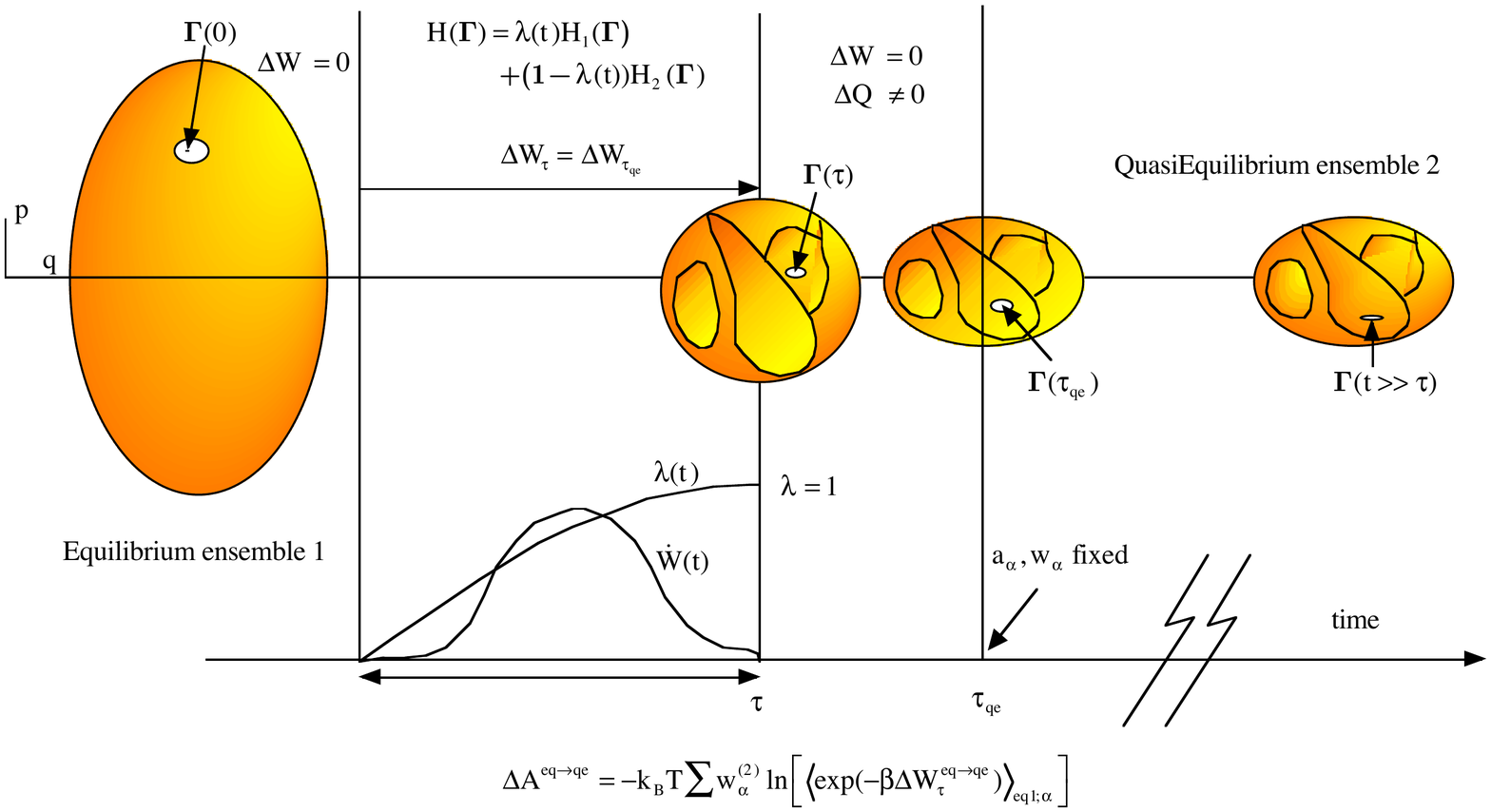}

\caption{\label{fig:schematic}A schematic diagram showing how the phase space
density (represented by shading) and the location of a phase space
volumes centred at ${\bf \Gamma}(0)$ evolve with time from an equilibrium
state at $time=0$ through a period, $\tau$ where the Hamiltonian
is changing with time, and then as the system relaxes to a quasiequilibrium
state at long times.}

\end{figure}

Following reference \citep{Williams-JCP-07}, for $t=\tau_{qe}$ we
can write the distribution function of a single occupied domain, $D_{\alpha}$,
as

\begin{equation}
f_{\alpha}(\boldsymbol{\Gamma})=\frac{s_{\alpha}(\mathbf{\Gamma})\exp[-\beta H_{0}(\boldsymbol{\Gamma})]}{\int_{D_{\alpha}}d\boldsymbol{\Gamma}\exp[-\beta H_{0}(\boldsymbol{\Gamma})]}=\frac{s_{\alpha}(\mathbf{\Gamma})\exp[-\beta H_{0}(\boldsymbol{\Gamma})]}{\int d\boldsymbol{\Gamma}s_{\alpha}(\mathbf{\Gamma})\exp[-\beta H_{0}(\boldsymbol{\Gamma})]}=\frac{s_{\alpha}(\mathbf{\Gamma})\exp[-\beta H_{0}(\boldsymbol{\Gamma})]}{Z_{\alpha}}\label{domain-distribution}\end{equation}
where the switch, $s_{\alpha}(\boldsymbol{\Gamma})$, is equal to
unity when $\boldsymbol{\Gamma}\in D_{\alpha}$ and zero otherwise,
and \begin{equation}
Z_{\alpha}=\int d\boldsymbol{\Gamma}s_{\alpha}(\mathbf{\Gamma})\exp[-\beta H_{0}(\boldsymbol{\Gamma})].\label{eq:Z_alpha}\end{equation}
We note that $f_{\alpha}(\boldsymbol{\Gamma})$ is the phase space
density at ${\bf \Gamma}$ normalised over $D_{\alpha}$ \emph{only}.
Only in the case of an equilibrium state will $f_{\alpha}(\boldsymbol{\Gamma})=f(\boldsymbol{\Gamma})$
for all ${\bf \Gamma}$. We now write the distribution function for
the quasiequilibrium ensemble,

\begin{equation}
f_{qe}(\boldsymbol{\Gamma})=\frac{\sum_{\alpha=1}^{N_{D}}a_{\alpha}\, s_{\alpha}(\mathbf{\Gamma})\exp[-\beta H_{0}(\boldsymbol{\Gamma})]}{\sum_{\gamma=1}^{N_{D}}a_{\gamma}\int_{D_{\gamma}}d\boldsymbol{\Gamma}\exp[-\beta H_{0}(\boldsymbol{\Gamma})]}=\frac{\sum_{\alpha=1}^{N_{D}}a_{\alpha}\, s_{\alpha}(\mathbf{\Gamma})\exp[-\beta H_{0}(\boldsymbol{\Gamma})]}{Z_{Z}}=\frac{\sum_{\alpha=1}^{N_{D}}a_{\alpha}\, f_{\alpha}(\mathbf{\Gamma})Z_{\alpha}}{Z_{Z}},\label{dist_fn_quasiequilibrium}\end{equation}
where the partition function \begin{equation}
Z_{Z}\equiv\sum_{\alpha=1}^{N_{D}}a_{\alpha}\int_{D_{\alpha}}d\boldsymbol{\Gamma}\exp[-\beta H_{0}(\boldsymbol{\Gamma})].\label{eq:Z_z}\end{equation}
The value of $a_{\alpha}$ gives the contribution of the domain $\alpha$
to the partition function, relative to its contribution in an equilibrium
state. If the domain $\alpha$ is unoccupied, $a_{\alpha}=0$. If
we consider an equilibrium distribution of states that is arbitrarily
partitioned into domains then we see that if the same partition function
is to be obtained by summing over the arbitrary domains as was obtained
without partitioning, then 

\begin{equation}
a_{\alpha}=1,\;\forall\;\alpha.\label{eq:equilibrium condition}\end{equation}
This in turn implies the quasiequilibrium normalisation condition:

\begin{equation}
\sum_{\alpha=1}^{N_{D}}a_{\alpha}=N_{D}.\label{eq:a-normalization}\end{equation}
We can define a partition function weighted free energy of an ensemble
of glass samples, $A_{Z}$, as 

\[
\exp[-\beta A_{Z}]\equiv Z_{Z}=\sum_{\alpha=1}^{N_{D}}a_{\alpha}\int_{D_{\alpha}}d\boldsymbol{\Gamma}\exp[-\beta H_{0}(\boldsymbol{\Gamma})]\]

\[
=\sum_{\alpha=1}^{N_{D}}a_{\alpha}\int d\boldsymbol{\Gamma}s_{\alpha}(\mathbf{\Gamma})\exp[-\beta H_{0}(\boldsymbol{\Gamma})]\]

\begin{equation}
=\sum_{\alpha=1}^{N_{D}}a_{\alpha}Z_{\alpha}=\sum_{\alpha=1}^{N_{D}}a_{\alpha}\exp[-\beta A_{\alpha}]\label{quasiHelmholtz_glass}\end{equation}
where the local domain free energy is $A_{\alpha}=-k_{B}T\ln Z_{\alpha}$.
For ease of reference we will refer to $A_{Z}$ as the quasi-Helmholtz
free energy. We will show later that (except at equilibrium) this
free energy is \emph{not} the Helmholtz free energy. 

Replacing the equilibrium distribution function in the definition
Eq. (\ref{generalised work_distfns}) with the quasiequilibrium distribution
functions, Eq. (\ref{dist_fn_quasiequilibrium}), and using $Z_{Z}$
for the partition functions, it is straightforward to show that for
an ensemble of glass samples,

\[
\exp[\Delta X_{Z,\tau_{qe}}(\mathbf{\Gamma})]=\frac{f_{qe}^{(1)}(\mathbf{\Gamma})\left\Vert \partial\mathbf{\Gamma}/\partial\mathbf{\mathbf{\Gamma}}(\tau_{qe})\right\Vert Z_{Z}^{(1)}}{f_{qe}^{(2)}(\mathbf{\Gamma}(\tau_{qe}))Z_{Z}^{(2)}}\]

\begin{eqnarray}
 & = & \frac{\sum_{\alpha=1}^{N_{D}}a_{\alpha}^{(1)}\, s_{\alpha}^{(1)}(\mathbf{\Gamma})\exp[-\beta H_{0}(\boldsymbol{\Gamma})]\left\Vert \partial\mathbf{\Gamma}/\partial\mathbf{\mathbf{\Gamma}}(\tau_{qe})\right\Vert }{\sum_{\gamma=1}^{N_{D}}a_{\gamma}^{(2)}\, s_{\gamma}^{(2)}(\mathbf{\Gamma}(\tau_{qe}))\exp[-\beta H_{0}(\boldsymbol{\Gamma}(\tau_{qe}))]}\nonumber \\
 & = & \exp\left[\beta(H_{0}(\boldsymbol{\Gamma}(\tau_{qe}))-H_{0}(\boldsymbol{\Gamma})\right]\left\Vert \partial\mathbf{\Gamma}/\partial\mathbf{\mathbf{\Gamma}}(\tau_{qe})\right\Vert \frac{\sum_{\alpha=1}^{N_{D}^{(1)}}a_{\alpha}^{(1)}\, s_{\alpha}^{(1)}(\mathbf{\Gamma})}{\sum_{\gamma=1}^{N_{D}^{(2)}}a_{\gamma}^{(2)}\, s_{\gamma}^{(2)}(\mathbf{\Gamma}(\tau_{g}))}\nonumber \\
 & = & \exp\left[\beta\Delta W_{\tau_{qe}}({\bf \Gamma})\right]\frac{\sum_{\alpha=1}^{N_{D}^{(1)}}a_{\alpha}^{(1)}\, s_{\alpha}^{(1)}(\mathbf{\Gamma})}{\sum_{\gamma=1}^{N_{D}^{(2)}}a_{\gamma}^{(2)}\, s_{\gamma}^{(2)}(\mathbf{\Gamma}(\tau_{qe}))}.\label{generalised work_qedf}\end{eqnarray}
We note that $\Delta W_{\tau}(\boldsymbol{\Gamma})=\Delta W_{\tau_{qe}}(\boldsymbol{\Gamma})$
since no work is done during the relaxation period $\tau<t<\tau_{qe}$,
and $\Delta W_{\tau}(\boldsymbol{\Gamma}(0))$ is the work given by
Eq. (\ref{work}). In Eq. (\ref{generalised work_qedf}) $a_{\alpha}^{(1)},\: a_{\gamma}^{(2)}$
are the weights for the relaxed quasiequilibrium state which, by choice
of $\tau_{qe}$, are the domain weights at that same finite time.
Note at this same time the intra-domain weights may not yet be Boltzmann.
Therefore,

\begin{equation}
\Delta X_{Z,\tau_{qe}}(\boldsymbol{\Gamma})=\beta\Delta W_{\tau}(\boldsymbol{\Gamma})+\ln\left[\sum_{\alpha=1}^{N_{D}^{(1)}}a_{\alpha}^{(1)}s_{\alpha}^{(1)}(\mathbf{\Gamma})\right]-\ln\left[\sum_{\gamma=1}^{N_{D}^{(2)}}a_{\gamma}^{(2)}s_{\gamma}^{(2)}(\mathbf{\Gamma}(\tau_{qe}))\right].\label{work_glass}\end{equation}
The expression Eq. (\ref{generalised work_qedf}) was obtained from
Eq. (\ref{generalised work_distfns}) and therefore $\Delta X_{Z,\tau_{qe}}(\boldsymbol{\Gamma}(0))$
also satisfies a modified version of Eq. (\ref{generalised Jarzynski}):

\begin{equation}
\left\langle \exp(-\Delta X_{Z,\tau_{qe}})\right\rangle _{1}=\frac{Z_{Z}^{(2)}}{Z_{Z}^{(1)}}=\exp[-\beta\Delta A_{Z}].\label{Nonequilibrium AZ expression}\end{equation}
Due to its resemblance to the GJE, Eq. (\ref{generalised Jarzynski}),
we refer to this as the GJE for quasiequilibrium ensembles, or simply
the quasiequilibrium GJE. In the derivation of this relationship we
have assumed:\resume{enumerate}

\item The occupancy of the domains in the final state at $t=\tau_{qe}$
is the same occupancy as in the relaxed quasiequilibrium state.\label{enu:constant occupancy}
\item For every $\mathbf{\Gamma}$ where $f_{qe}^{(1)}(\mathbf{\Gamma})\neq0$
we require that $f_{qe}^{(2)}(\mathbf{\Gamma}(\tau_{qe}))\neq0$ and
vice versa. Translating this into domain weights, if $s_{\alpha}^{(1)}(\mathbf{\Gamma})\neq0$
then we require that $s_{\beta}^{(2)}(\mathbf{\Gamma}(\tau_{qe}))\neq0$,
and vice versa. \label{enu:nonzero domain weights at tqe}
\end{enumerate}
The first assumption arises because although $\Delta W_{\tau}(\boldsymbol{\Gamma}(0))$
does not change beyond $t=\tau$, the weights $a_{\alpha}^{(2)}$
and $s_{\alpha}^{(2)}(\mathbf{\Gamma}(t))$ do continue to change
for $t>\tau$. This is very different to the usual circumstance for
ergodic systems. We also note that a necessary condition for the intra-domain
weights to be Boltzmann is that they relax on a time scale which is
much shorter than that for the inter-domain weights. This implies
that although the inter-domain weights do not significantly change
for $t\geq\tau_{qe}$, at the time $\tau_{qe}$ the intra-domain weights
may not yet be Boltzmann. 

For an aged glass where the observable rate of change for any macroscopic
property relative to the relaxation time of the intra-domain weights
approaches infinity, we have given a proof that if the phase space
domains are robust with respect to small changes in macroscopic parameters,
the phase space distribution within any domain is distributed in a
Maxwell-Boltzmann distribution. Such distributions always satisfy
assumption \ref{enu:constant occupancy} above \citep{Williams-JCP-07}.

We note that if there is only one occupied domain, (e.g. $a_{1}=1$
and $a_{2},\ldots,\: a_{N_{D}}=0$) then $A_{Z}=A_{1}-k_{B}T\ln(1)=A_{1}$
where $A_{1}$ is defined as in Eq. (\ref{quasiHelmholtz_glass}).

If the initial and final states are both at equilibrium, $\Delta A_{Z}=\Delta A$
since $a_{i}=1\:\forall\: i$. However, in general this is not the
case. In order to determine the relationship between the Helmholtz
free energy difference between quasiequilibrium states and $\Delta A_{Z}$,
we consider its thermodynamic definition. The Gibbs expression for
the entropy is\begin{equation}
S\equiv-k_{B}\int d\boldsymbol{\Gamma}\: f(\boldsymbol{\Gamma})\ln[f(\boldsymbol{\Gamma})],\label{entropy}\end{equation}
where the integral is over all space. Once the nonequilibrium process
has finished, and the system slowly relaxes towards equilibrium ($t>\tau$),
we have every expectation that Eq. (\ref{entropy}) obeys the second
law inequality, that is\begin{equation}
T\frac{dS}{dt}\geq\frac{d\left\langle Q\right\rangle }{dt}.\label{SLI}\end{equation}
When the transformation process finishes no more work is done on the
system and the only way the average energy can change is through the
transfer of heat, in our case through thermostats. Given the fundamental
thermodynamic relation for the Helmholtz free energy,\begin{equation}
A\equiv\left\langle H_{0}\right\rangle -TS,\label{A-thermo-relation}\end{equation}
and the equation for the entropy, Eq. (\ref{entropy}), the Helmholtz
free energy is uniquely defined. In reference \citep{Williams-JCP-07},
it was shown that quasiequilibrium states can be treated using standard
macroscopic thermodynamics. So by use of the distribution function,
Eq. (\ref{dist_fn_quasiequilibrium}), we obtain the Helmholtz free
energy of a quasiequilibrium state \[
A_{qe}=\left\langle H_{0}\right\rangle _{qe}+k_{B}T\left\langle \ln[f_{qe}(\boldsymbol{\Gamma})]\right\rangle _{qe}\]

\[
=\left\langle H_{0}\right\rangle _{qe}+k_{B}T\left\langle \ln\left[\sum_{\alpha=1}^{N_{D}}a_{\alpha}s_{\alpha}(\boldsymbol{\Gamma})exp[-\beta H_{0}(\boldsymbol{\Gamma})]/Z_{Z}\right]\right\rangle _{qe}\]

\[
=\left\langle H_{0}\right\rangle _{qe}-k_{B}T\left\langle \ln\left[\sum_{\alpha=1}^{N_{D}}s_{\alpha}(\boldsymbol{\Gamma})a_{\alpha}\right]\right\rangle _{qe}+k_{B}T\left\langle -\beta H_{0}(\boldsymbol{\Gamma})-\ln[Z_{Z}]\right\rangle _{qe}\]
\begin{equation}
=-k_{B}T\left\langle \ln\left[\sum_{\alpha=1}^{N_{D}}s_{\alpha}(\boldsymbol{\Gamma})a_{\alpha}\right]\right\rangle _{qe}+A_{Z}\label{Helmholtz in terms of AZ}\end{equation}
where the notation $\left\langle B(\mathbf{\Gamma})\right\rangle _{qe}\equiv\int_{D}d\Gamma\, f_{qe}(\mathbf{\Gamma})B(\mathbf{\Gamma})$
where $D$ is all the available phase space in the glass state. Therefore,
calculation of $A_{Z}$ which is an ensemble average of $\Delta X_{Z,\tau_{qe}}$
calculated along nonequilibrium trajectories, and use of Eq. (\ref{Helmholtz in terms of AZ})
allows the Helmholtz free energy of a quasiequilibrium state to be
determined:

\begin{equation}
\Delta A_{qe}=\Delta A_{Z}+k_{B}T\left\langle \ln\left[\sum_{\alpha=1}^{N_{D}^{(1)}}a_{\alpha}^{(1)}s_{\alpha}^{(1)}(\mathbf{\Gamma})\right]\right\rangle -k_{B}T\left\langle ln\left[\sum_{\gamma=1}^{N_{D}^{(2)}}a_{\gamma}^{(2)}s_{\gamma}^{(2)}(\mathbf{\Gamma}(\tau_{qe}))\right]\right\rangle .\label{quasi-Helmholtz 1}\end{equation}

In Eq. (27) of reference \onlinecite{Williams-JCP-07}, it was shown
(for Gibbs free energies rather than Helmholtz) that the free energy
was minimised when all the domain weights were Boltzmann, that is
$\int_{D_{\alpha}}d{\bf \Gamma}\frac{a_{\alpha}\exp(-\beta H_{0}({\bf \Gamma}))}{Z_{Z}}=\int_{D_{\alpha}}d{\bf \Gamma}\frac{\exp(-\beta H_{0}({\bf \Gamma}))}{Z}$.
This implies that $A(a_{1},a_{2}\ldots a_{N_{D}})$ is minimised when
$a_{i}=1\:\:\forall\:\: i$, which coincides with thermodynamic equilibrium.
We make the standard observation of macroscopic thermodynamics that
when the system is not acted on externally Eqs. (\ref{SLI}) \& (\ref{A-thermo-relation})
give $dA/dt\leq0$. Thus we have proved the following: as the system's
distribution function moves towards the equilibrium state, which is
at the point $a_{\alpha}=1\:\:\forall\:\:\alpha$, Eq. (\ref{SLI})
is obeyed. By \emph{towards} we mean the direction the system is moving
has a component in the direction given by $-\nabla A_{qe}+[(\nabla A_{qe}\cdot\nabla g)/(\nabla g\cdot\nabla g)]\nabla g$
where $g=\sum_{\alpha=1}^{N_{D}}a_{\alpha}-N_{D}$ and $\nabla$ acts
on the $N_{D}$ dimensional space given by the coordinate set $a_{\alpha}$.
In contrast $A_{Z}$ might not be a minimum in equilibrium.

\subsection{Quasiequilibrium free energies from a weighted sum of local domain
free energies.}

A nonequilibrium free energy relation can also be obtained for systems
that are quenched from an equilibrium state to a quasiequilibrium
state by considering the average of $\exp(-\beta\Delta W_{\tau})$
over trajectories that are in domain $D_{\alpha}$ of a quasiequilibrium
system at time $\tau$. These can then be combined to obtain a difference
in free energy of the initial equilibrium state and the final quasiequilibrium
state. In this case it is convenient to work using a weighted sum
of distributions that are normalised over sub-domains as introduced
by Williams and Evans \citep{Williams-JCP-07}:\begin{eqnarray}
f_{qe}(\mathbf{\Gamma}) & = & \sum_{\alpha=1}^{N_{D}}w_{\alpha}f_{\alpha}(\boldsymbol{\Gamma})\nonumber \\
 & = & \sum_{\alpha=1}^{N_{D}}\frac{w_{\alpha}s_{\alpha}(\mathbf{\Gamma})\exp[-\beta H_{0}(\boldsymbol{\Gamma})]}{\int_{D_{\alpha}}d\boldsymbol{\Gamma}\exp[-\beta H_{0}(\boldsymbol{\Gamma})]}.\label{sum-distribution-WE}\end{eqnarray}
Here $N_{D}$ is the number of domains, and $w_{\alpha}$ represents
the relative weights of these domains under the constraint $\sum_{\alpha=1}^{N_{D}}w_{\alpha}=1$,
see reference \onlinecite{Williams-JCP-07}. As above, we assume conditions
\ref{enu:ergoconsistGenWork1}-\ref{enu:ErgoConsistGJE} hold. The
relationship between $w_{\alpha}$ and $a_{\alpha}$ can be obtained
by considering Eqs. (\ref{dist_fn_quasiequilibrium}) \& (\ref{sum-distribution-WE})
and is given by:\begin{equation}
w_{\alpha}=\frac{a_{\alpha}\int_{D_{\alpha}}d\boldsymbol{\Gamma}\exp[-\beta H_{0}(\boldsymbol{\Gamma})]}{\sum_{\gamma=1}^{N_{D}}a_{\gamma}\int_{D_{\gamma}}d\boldsymbol{\Gamma}\exp[-\beta H_{0}(\boldsymbol{\Gamma})]}=\frac{a_{\alpha}Z_{\alpha}}{Z_{Z}}.\label{weights relationship}\end{equation}

In order to develop free energy relations, we consider two possibilities:
in the first case we monitor the work as the system is quenched from
an initial equilibrium state to a quasiequilibrium state by varying
$H_{0}$ over a period $0<t\le\tau$; in the second case we consider
the reverse process where the work is monitored as a prepared quasiequilibrium
system is relaxed towards the ergodic equilibrium state by varying
$H_{0}$ over a period $0<t\le\tau$ using the reverse protocol. In
the first case we can use the relationships between the work and equilibrium
canonical distribution functions, Eqs. (\ref{generalised Jarzynski})
\& (\ref{work}), to show for any domain, $D_{\alpha}$, of a quasiequilibrium
state,

\[
\left\langle s_{\alpha}(\mathbf{\Gamma}(\tau_{qe}))\exp(-\beta\Delta W_{\tau}^{eq\rightarrow qe}(\mathbf{\Gamma)})\right\rangle _{eq1}=\int d\mathbf{\Gamma}f_{eq}^{(1)}(\mathbf{\Gamma})\frac{s_{\alpha}(\mathbf{\Gamma}(\tau_{qe}))f_{eq}^{(2)}(\mathbf{\Gamma}(\tau_{qe}))\left\Vert \partial\mathbf{\Gamma}(\tau_{qe})/\partial\mathbf{\Gamma}\right\Vert Z^{(2)}}{f_{eq}^{(1)}(\mathbf{\Gamma})Z^{(1)}}\]
\begin{equation}
=\int d\mathbf{\mathbf{\Gamma}}(\tau_{qe})\frac{s_{\alpha}(\mathbf{\Gamma}(\tau_{qe}))f_{eq}^{(2)}(\mathbf{\Gamma}(\tau_{qe}))Z^{(2)}}{Z^{(1)}}=\frac{Z_{\alpha}^{(2)}}{Z^{(1)}}=\exp[-\beta\Delta A_{2\alpha,1}]\label{quasi <exp-beta.W>}\end{equation}
where we have used the fact that $\Delta W_{\tau}^{eq\rightarrow qe}({\bf \Gamma})=\Delta W_{\tau_{qe}}^{eq\rightarrow qe}({\bf \Gamma})$
and $\Delta A_{2\alpha,1}=A_{\alpha}^{(2)}-A^{(1)}$. Note that only
trajectories that are in $D_{\alpha}$ at time $\tau_{qe}$ will have
a non-zero contribution to the ensemble average on the left. Using
this and the Schr\"odinger-Heisenberg equivalence for phase space
averages (see Section 3.3 of reference \onlinecite{evans-and-morriss-book})
we may write $\left\langle s_{\alpha}(\mathbf{\Gamma}(\tau_{qe}))\right\rangle _{eq1}=\left\langle s_{\alpha}(\mathbf{\Gamma})\right\rangle _{qe}=w_{\alpha}$.
By only averaging over trajectories that are in $D_{\alpha}$ at time
$\tau$, we form a conditional ensemble average $\left\langle B(\mathbf{\Gamma)}\right\rangle _{eq1;\alpha_{qe}}\equiv\frac{\int d\mathbf{\Gamma}B({\bf \Gamma})f_{eq}^{(1)}(\mathbf{\Gamma})s_{\alpha}(\mathbf{\Gamma}(\tau_{qe}))}{\int d\mathbf{\Gamma}f_{eq}^{(1)}(\mathbf{\Gamma})s_{\alpha}(\mathbf{\Gamma}(\tau_{qe}))}=\frac{\left\langle s_{\alpha}(\mathbf{\Gamma}(\tau_{qe}))B({\bf \Gamma})\right\rangle _{eq1}}{\left\langle s_{\alpha}(\mathbf{\Gamma}(\tau_{qe}))\right\rangle _{eq1}}$,
so using Eq. (\ref{quasi <exp-beta.W>}) we can write:

\begin{equation}
\left\langle \exp(-\beta\Delta W_{\tau}^{eq\rightarrow qe}(\mathbf{\Gamma)})\right\rangle _{eq1;\alpha_{qe}}\equiv\frac{\left\langle s_{\alpha}(\mathbf{\Gamma}(\tau_{qe}))\exp(-\beta\Delta W_{\tau}^{eq\rightarrow qe}(\mathbf{\Gamma)})\right\rangle _{eq1}}{\left\langle s_{\alpha}(\mathbf{\Gamma}(\tau_{qe}))\right\rangle _{eq1}}=\frac{1}{w_{\alpha}}\exp[-\beta\Delta A_{2\alpha,1}].\label{conditional quasi <exp-beta.W>}\end{equation}
We emphasise that by $\left\langle \ldots\right\rangle _{eq1;\alpha_{qe}}$
we imply that the condition is on the domain that the trajectory is
in when it reaches the quasiequilibrium state, and not on the domain
in which it starts. 

Note that in Eqs. (\ref{work_glass}) \& (\ref{quasi <exp-beta.W>}),
we have assumed that $w_{\alpha}$ and $a_{\alpha}$ do not vary with
time (they remain equal to their value at $t=\tau_{qe}$). Of course
this does not allow for the extremely slow relaxation to the final
equilibrium state that might occur, on a much larger timescale. Therefore
it is more accurate to say that they are constant on the accessible
timescales. 

Conditions \ref{enu:ergoconsistGenWork1} \& \ref{enu:constant occupancy}
specified above for the quasiequilibrium GJE are also required in
this case. However, as discussed above, these restrictions are not
likely to be of any significance for cases of practical interest,
certainly not from the energy landscape point of view \citep{Wales-Book}.

From Eq. (\ref{conditional quasi <exp-beta.W>}), and the fact that
the Helmholtz free energy $A=\sum_{\alpha=1}^{N_{D}}[w_{\alpha}A_{\alpha}+k_{B}Tw_{\alpha}\ln(w_{\alpha})]$
(see Eq. (25) of reference \onlinecite{Williams-JCP-07}), we obtain, 

\begin{equation}
\Delta A^{eq\rightarrow qe}=-k_{B}T\sum_{\alpha=1}^{N_{D}}w_{\alpha}^{(2)}\ln\left[\left\langle \exp(-\beta\Delta W_{\tau}^{eq\rightarrow qe}(\mathbf{\Gamma)})\right\rangle _{eq1;\alpha}\right]\label{NEFERQEqm}\end{equation}
where $\Delta A^{eq\rightarrow qe}=A_{qe}^{(2)}-A^{(1)}$ is the difference
between the initial equilibrium Helmholtz free energy $A^{(1)}$ and
the Helmholtz free energy of the nonergodic quasiequilibrium state
$A_{qe}^{(2)}$.

Alternatively, we can develop an expression for the ensemble average
over a single domain of the quasiequilibrium state, by considering
the reverse process when the initial state (state (2)) is a relaxed
quasiequilibrium state and the final state (state (1)) is an ergodic
equilibrium state. Since each domain in the quasiequilibrium state
is locally canonical, the same arguments in Section A can be used
to show that for any trajectory starting at ${\bf \Gamma}$ in domain
$D_{\alpha}$ and subject to a change in $H_{0}$ over a period $0<t\le\tau$,
(rather than the longer period $0<t\le\tau_{qe}$)

\begin{equation}
\exp(\beta\Delta W_{\tau}^{qe\rightarrow eq}(\mathbf{\Gamma}))=\frac{f_{\alpha}^{(2)}(\mathbf{\Gamma})\left\Vert \partial{\bf \Gamma}/\partial{\bf \Gamma}(\tau)\right\Vert Z_{\alpha}^{(2)}}{f_{eq}^{(1)}(\mathbf{\Gamma}(\tau))Z_{}^{(1)}}.\end{equation}
The ensemble average over domain $D_{\alpha}$ is given by

\begin{eqnarray}
\left\langle \exp(-\beta\Delta W_{\tau}^{qe\rightarrow eq}(\mathbf{\Gamma}))\right\rangle _{\alpha_{qe}} & = & \int d\mathbf{\Gamma}s_{\alpha}({\bf \Gamma})f_{\alpha}^{(2)}(\mathbf{\Gamma})\frac{f_{eq}^{(1)}(\mathbf{\Gamma}(\tau))\left\Vert \partial\mathbf{\Gamma}(\tau)/\partial\mathbf{\Gamma}\right\Vert Z^{(1)}}{f_{\alpha}^{(2)}(\mathbf{\Gamma})Z_{\alpha}^{(2)}}\nonumber \\
 & = & \frac{Z^{(1)}}{Z_{\alpha}^{(2)}}\int d\mathbf{\Gamma}(\tau)\: s_{\alpha}({\bf \Gamma})f_{eq}^{(1)}(\mathbf{\Gamma}(\tau))\nonumber \\
 & = & \exp\left[-\beta(A^{(1)}-A_{\alpha}^{(2)}-k_{B}Tw_{\alpha})\right]\label{conditional-backward-glass-JE}\end{eqnarray}
where the Schr\"odinger-Heisenberg equivalence for phase space averages
is used to give $\int d\mathbf{\Gamma}(\tau)\: s_{\alpha}({\bf \Gamma})f_{eq}^{(1)}(\mathbf{\Gamma}(\tau))=\left\langle s_{\alpha}(\mathbf{\Gamma})\right\rangle _{qe}=w_{\alpha}$
and obtain the final equality. As above, we can then show that 

\begin{equation}
\Delta A^{qe\rightarrow eq}=-k_{B}T\sum_{\alpha=1}^{N_{D}}w_{\alpha}^{(2)}\ln\left[\left\langle \exp(-\beta\Delta W_{\tau}^{qe\rightarrow eq}(\mathbf{\Gamma)})\right\rangle _{\alpha_{qe}}\right].\label{backward-glass-JE}\end{equation}

In this section and Section B, we have described three approaches
for determination of the free energy of the quasiequilibrium state
that involve exponential averages of nonequilibrium path integrals,
in the same way the Jarzynski Equality is applied to equilibrium states.
In the next section we will consider practical issues regarding their
application.

\section{Model and Simulation Details\label{sec-Mod-and-Sim-det}}

In order to examine the ability of the Jarzynski Equality and the
new free energy expressions Eqs. (\ref{Helmholtz in terms of AZ})
\& (\ref{backward-glass-JE}) to probe the free energy of a quasiequilibrium
state, we consider a simple model of a system which may be quenched
into a glass state. We employ a dynamical model originally developed
by Hoover and coworkers \citep{Hoover-CMP-2005,Hoover-MS-2007} for
other purposes. This simple dynamical system is ergodic and mixing
and samples phase space canonically despite there only being a single
particle in a one dimensional Cartesian space. To achieve this two
Nos\'e-Hoover thermostats are employed giving the following equations
of motion\begin{eqnarray}
\dot{q} & = & \frac{p}{m}\nonumber \\
\dot{p} & = & F(q)-\zeta_{1}p-\zeta_{3}p^{3}\label{EOM}\\
\dot{\zeta}_{1} & = & \left(\beta\frac{p^{2}}{m}-1\right)/\tau_{1}^{2}\nonumber \\
\dot{\zeta}_{3} & = & \left(\beta\frac{p^{4}}{m}-3p^{2}\right)/\tau_{3}^{2},\nonumber \end{eqnarray}
where $q$ is the particle's position, $p$ is its momentum, $\beta=1/(k_{B}T$)
where $T$ is the average temperature regulated by the two thermostats
and $F(q)=-d\Phi(q)/dq$ is the force acting on the particle. The
variables $\tau_{1}$ and $\tau_{3}$ are time constants for the thermostat's
feedback mechanism. 

Using the Liouville theorem, the distribution function of this system
can be derived, and is given by \citep{Hoover-CMP-2005},

\begin{equation}
f\left(q,p,\zeta_{1},\zeta_{3}\right)=\frac{\tau_{1}\tau_{3}}{(2\pi)^{3/2}(mk_{B}T)^{1/2}}\frac{\exp(-\beta H_{e}(q,p,\zeta_{1},\zeta_{3})).}{\int_{-\infty}^{\infty}dq^{\prime}\exp(-\beta\Phi(q^{\prime})).}\label{distribution function}\end{equation}

\begin{figure}
\resizebox{8.5cm}{!}{\includegraphics{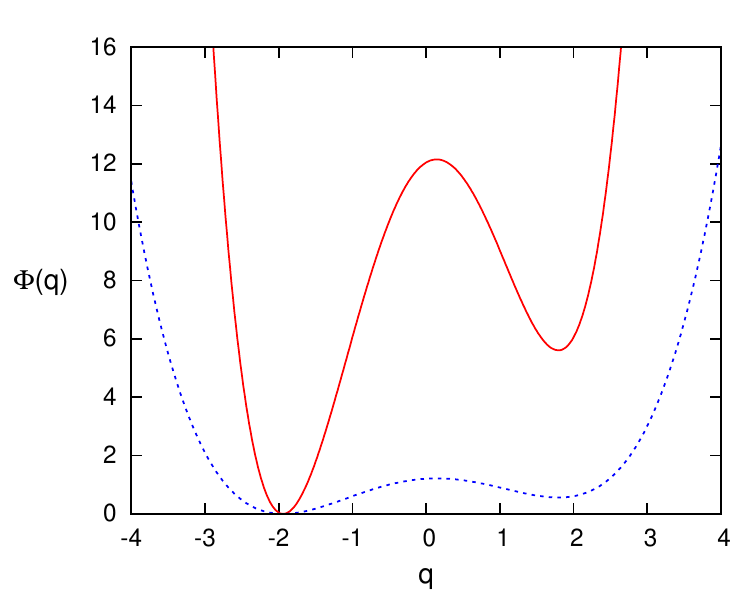}} 

\caption{\label{fig:potential}Potential energy function, defined by Eq. (\ref{potential})
with the blue dashed line corresponding to $\epsilon=0.1$ and the
red solid line corresponding to $\epsilon=1$.}

\end{figure}

\noindent Here $H_{e}(q,p,\zeta_{1},\zeta_{3})=H_{0}(q,p)+\frac{1}{2}k_{B}T(\tau_{1}^{2}\zeta_{1}^{2}+\tau_{3}^{2}\zeta_{3}^{2})=\Phi(q)+\frac{1}{2}[p^{2}/m+k_{B}T(\tau_{1}^{2}\zeta_{1}^{2}+\tau_{3}^{2}\zeta_{3}^{2})]$
where $H_{0}$ is the Hamiltonian and internal energy of the unthermostatted
oscillator. The partition function is

\noindent \begin{equation}
Z_{e}=\frac{(2\pi)^{3/2}(mk_{B}T)^{1/2}\int_{-\infty}^{\infty}dq^{\prime}\exp(-\beta\Phi(q^{\prime}))}{\tau_{1}\tau_{3}}=\frac{2\pi}{\tau_{1}\tau_{3}}Z,\label{oscillator partition function}\end{equation}
where $Z_{e}$ is the partition function in the extended phase space.
We use a double well potential to form a simple model of a glass.
This forms a very simplistic representation of the complicated energy
landscape of a real glass. It features a local minimum that can be
separated from the global minimum on quenching. The potential, shown
in Fig. \ref{fig:potential}, is given by the equation\begin{equation}
\Phi(q)=\epsilon\left(b_{0}+b_{1}q/\sigma+b_{2}q^{2}/\sigma^{2}+b_{4}q^{4}/\sigma^{4}\right),\label{potential}\end{equation}
where $b_{0}=12.04541125$, $b_{1}=1.5$, $b_{2}=-5.25$ and $b_{4}=0.75$.
We use reduced units throughout this section, where the length unit
is $\sigma$, the mass unit is $m$ and the energy unit is $k_{B}T$,
resulting in the time unit $\sigma\sqrt{m/k_{B}T}$. The thermostat
time constants have fixed values of $\tau_{1}=\tau_{3}=0.5$. The
potential has a global minimum of $\Phi=0$ at $q=1.79483214$, a
local minimum at $q=-1.9385372$ and a local maximum at $q=0.14370505$.
In Fig. \ref{fig:potential} the potential is plotted for two different
values of $\epsilon$ (in units of $k_{B}T$). When $\epsilon=0.1$,
the two local minima are separated by a barrier of the order of 1
(i.e. $k_{B}T$), a barrier that the system readily traverses. At
the lower temperature, when $\epsilon=1.0$, (i.e. $k_{B}T$) a system
in the local minimum is separated from the global minimum by a barrier
of the order of 6 (i.e. $6k_{B}T)$. This is a significant energy
barrier and the crossing of it constitutes a rare event. After quenching
a large ensemble of systems to this low temperature, the higher energy
local minimum is populated by a larger proportion of the ensemble
than it would be when fully equilibrated. Due to the high energy barrier
this situation persists for a long time.

In the quenching experiment considered above, the initial state can
be divided into two non-intersecting domains with $q<0.14370505$
and $q\ge0.14370505$, and both are occupied according to their equilibrium
distributions: $a_{1}=a_{2}=1$. For this state $Z_{Z}=Z$ is the
usual equilibrium partition function and $A_{Z}=A$ is the usual free
energy. In the final state we expect that, due to our selection of
the initial potential, after the quench the second domain ($q\ge0.14370505$)
will have an occupancy that is much higher than its Boltzmann occupancy
and the first will have a lower than Boltzmann occupancy. The equilibrated
occupancy for domain 2 is nearly zero so the change relative to the
Boltzmann level is very large. 

Here we have deliberately selected an initial distribution that leads
to $a_{2}\gg a_{1}$, so that we can clearly demonstrate the difference
in information provided by the Jarzynski Equality and the new free
energy expressions Eqs. (\ref{Helmholtz in terms of AZ}) \& (\ref{NEFERQEqm}).
The treatment described above can be readily applied to systems where
the values of the weights for more than one domain are significant.

In order to compare the Jarzynski Equality and the new relations,
ensembles of $10^{5}$ independent simulations were carried out starting
from an initial equilibrium ensemble with $\epsilon=0.1$ in Eq. (\ref{potential}).
At time $t=0$ the parameter $\epsilon$ was linearly increased to
a final value of $\epsilon=1$ at time $t=\tau=200$. A second set
of simulations was computed with duration of $\tau=2000$ rather than
$\tau=200$. These times were chosen to be sufficiently short that
a quasiequilibrium state develops, yet sufficiently long that within
the domains the distribution is sufficiently close to Boltzmann at
the end of the trajectory, that no further relaxation is required
to develop the quasiequilibrium distribution. That is, for our model
we can take $\tau=\tau_{qe}$. If a more rapid protocol was used,
this might not be the case and the system would need additional time
to relax before generating the quasiequilibrium distribution. The
probability distribution was then separated into two domains with
any configuration where the position of the particle was in the range
$-\infty<q<0.14370505$ being designated as in the first domain and
all other configurations $0.14370505\le q<\infty$ being designated
as in the second domain. Recall that the value $0.14370505$ is the
position of the local maximum in the potential Eq. (\ref{potential}).

\section{Results and Discussion}

\subsection{Distribution Functions}

\begin{figure}
\resizebox{8.5cm}{!}{\includegraphics{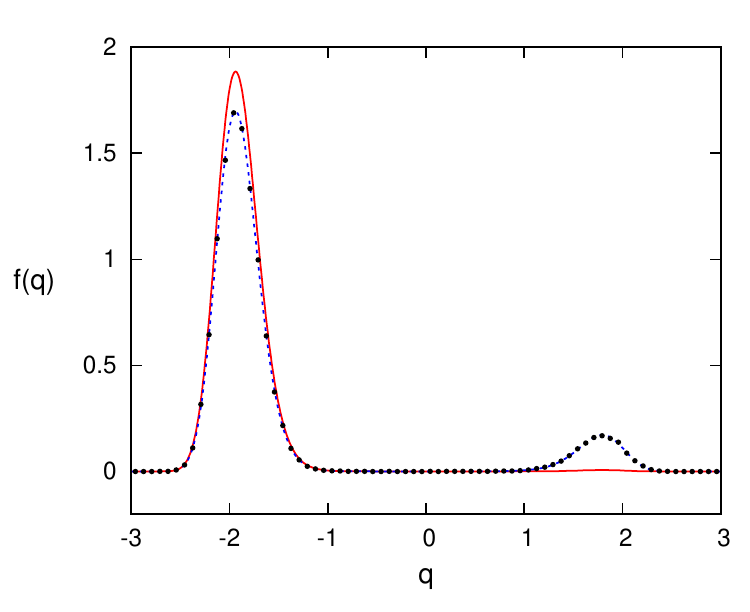}} 

\caption{\label{fig:distribution-functions}Distribution functions: the red
solid line is the equilibrium distribution function with $\epsilon=1.0$.
The black solid circles are numerical data for the distribution of
the quasiequilibrium state obtained by quenching the system over a
period $\tau=200$, and the blue dashed line is the best fit to the
numerical data, obtained by adjusting the single free parameter in
Eq. \ref{dist_fn_quasiequilibrium}.}

\end{figure}

The first point to be tested is whether Eq. (\ref{dist_fn_quasiequilibrium})
is able to accurately represent our simulations and to what degree
they are out of equilibrium at time $\tau=200$ with $\epsilon=1.0$.
In order to examine this, Fig. \ref{fig:distribution-functions} shows
$f(q)$ obtained from the equilibrium distribution function, Eq. (\ref{distribution function}),
and the distribution found in the simulations. Eq. (\ref{dist_fn_quasiequilibrium})
was then used to fit the data, and it was found that the best fit
was obtained with the single free parameter set to $a_{1}/a_{2}=0.03627$.
Using the normalisation condition gives $a_{1}=0.07000;\: a_{2}=1.930$.

It can be seen that Eq. (\ref{dist_fn_quasiequilibrium}) fits the
data very well despite $a_{1}$ and $a_{2}$ being very different,
which means the system is a long way from equilibrium by this measure.
In the vicinity of the local minima in the energy (at $q=1.79483214$),
the equilibrium distribution function can be seen to have a value
which is very close to zero. The observed quasiequilibrium distribution
function has a significant value here, approximately 25 times larger
than the equilibrium value. As time progresses, after the nonequilibrium
process has finished, this difference between the two distribution
functions follows a very slow exponential decay which will depend
on the barrier height in the potential \citep{Wales-Book} (see Fig.
\ref{fig:potential}) along with the choice of time constants $\tau_{1}$
and $\tau_{3}$ in Eq. (\ref{EOM}). We will not pursue the details
of this further in this paper.

The ergodic consistency conditions for the application of the Jarzynski
Equality and the quasiequilibrium GJE for the two quasiequilibrium
states are satisfied. Firstly there is only a very slow relaxation
of the weights after the transformations. The second condition namely
that if $s_{\alpha}^{(1)}(\mathbf{\Gamma})\neq0$ then we require
that $s_{\beta}^{(2)}(\mathbf{\Gamma}(\tau_{qe}))\neq0$, and vice
versa is also satisfied. Although it is hard to see in the figure
the final equilibrium distribution does have measurable density in
both domains and rather more obviously in the final quasiequilibrium
state there is density in both domains. There is also density in both
domains in the initial (ergodic) equilibrium distribution function.

\subsection{Standard Jarzynski Equality}

By employing the standard Jarzynski Equality, Eq. (\ref{JE}), we
are able to compute the difference in free energy between the equilibrium
states for $\epsilon=0.1$ and $\epsilon=1.0$. Of course, for the
simple model under consideration, the difference in free energy may
be readily computed using Eqs. (\ref{Helmholtz}), (\ref{oscillator partition function})
\& (\ref{potential}) for both values of $\epsilon$. Thus the partition
function was evaluated numerically, and a value of $\Delta A_{eq}=1.7219$
was obtained, where we add the subscript `eq' to indicate that both
states were at equilibrium. The value obtained from applying the Jarzynski
Equality, Eq. (\ref{JE}), to the ensemble of simulations of duration
$\tau=200$, was $\Delta A=1.7229$ and for the ensemble of duration
$\tau=2000$ it was $\Delta A=1.7231$. This strong agreement with
the equilibrium value in both cases gives a clear demonstration of
how the Jarzynski Equality gives the free energy difference between
the two equilibrated states. This might seem surprising. Despite the
fact that the period over which the work is measured is too short
to generate the final equilibrium state, the Jarzynski Equality refers
to the state that \emph{would} be reached after infinite relaxation
time. Therefore, although there may be a long-lasting glass state,
the Jarzynski Equality does not refer to that. 

Provided ergodic consistency is satisfied, the Jarzynski Equality
gives the difference between the initial equilibrium free energy and
the free energy of the final equilibrium state. However, if there
are phase space domains in the final state that are \emph{not sampled
at all,} the Jarzynski Equality gives the free energy difference between
the two \emph{equilibrium} states subject to the constraint that the
weights are zero in the excluded domains. The proof follows from a
simple gedanken experiment. Suppose that the reason why those excluded
domains are not sampled at all is that the potential energy is actually
infinite for all states in those excluded domains. Then clearly the
Jarzynski Equality gives the free energy difference between the two
equilibrium states defined using the modified potential energy function
for the final system. In this final equilibrium system all the nonzero
weights are Boltzmann.

In a real glassy state, the relaxation of the fluid is so slow that
the crystalline states are never reached and rather the glass remains
in a history dependent state for time scales beyond human experience.
Applying the Jarzynski Equality to glass forming systems where all
crystalline phases are never sampled, gives the free energy difference
between the initial equilibrium system and the {}``ideal'' or equilibrium
glass where the weights of all the glassy phases are Boltzmann - both
intra and inter domain weights. 

If the time over which $\epsilon$ varied was infinitely slow, the
system would be quasistatic and the process would be thermodynamically
reversible. The amount of work, Eq. (\ref{work}), done by every trajectory
would then be the same and equal to the change in Helmholtz free energy
and the instant the process finishes the system would be in equilibrium.
However in the final state, when $\epsilon$ is high, our system is
not ergodic on the time scale of our simulations. By ergodic we mean
that a single trajectory is able to sample a sufficient representation
of phase space to be accurately representative of the entire phase
space. Recalling the measured distribution function shown in Fig.
\ref{fig:distribution-functions}, this would require that $\epsilon$
does not change significantly during the time it takes a single trajectory
to sample a sufficient representation of the two peaks seen in the
figure. Here the occurrence of a trajectory crossing from one peak
to the other is a rare event and as $\epsilon$ is increased these
events become much rarer. Thus as $\epsilon$ increases, the minimal
time scale on which the system may not change significantly, in order
to obtain something representative of a quasistatic process, diverges. 

\begin{figure}
\resizebox{8.5cm}{!}{\includegraphics{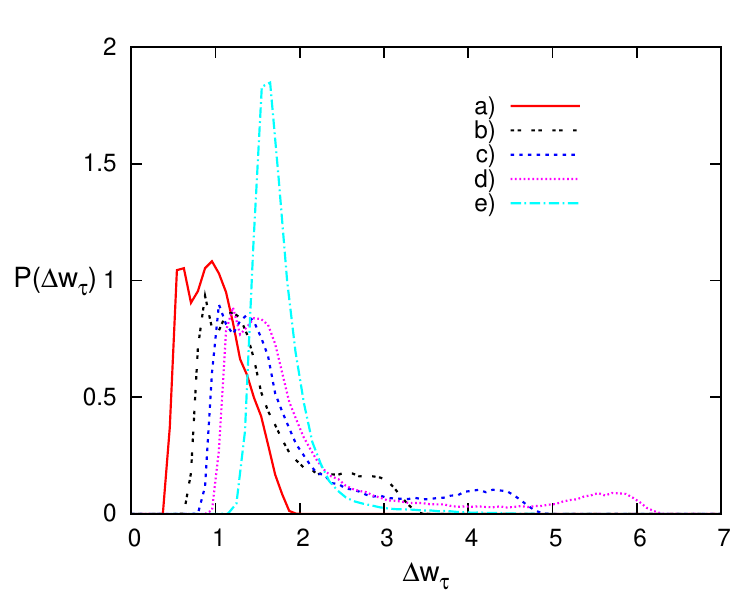}} 

\caption{\label{fig:work-distributions}The distribution functions for the
work done, Eq. (\ref{work}), for an ensemble of simulations where
the quench was carried out over period $\tau=200$. The distributions
are shown for the work done from when the quench states for times
of a) 50, b) 100, c) 150 and d) 200. Curve e) corresponds to distribution
of the work done for an ensemble of simulations when a quench was
carried out over a period $\tau=2000$. It shows the work done at
the end of the quench (at a time of 2000). }

\end{figure}

As the process is not thermodynamically reversible it is interesting
to consider the distribution functions for the work done by the trajectories
at different times, Fig. \ref{fig:work-distributions}. The distributions
are highly skewed towards large values of the work by trajectories
where the particle remains trapped in the local minimum of the potential
through the quench. By comparing the distribution of the more rapidly
and slowly quenched ensembles, $\tau=200$, and $\tau=2000$ respectively,
at the instant the quench finishes, $\epsilon=1$, we gain some insight
as to how the Jarzynski Equality works for this process. The distribution
for the slower quench is much sharper and less skewed, with only a
single peak, due to the process being significantly closer to the
quasistatic limit. Thus during the slower quench many trajectories
sample both the local minimum and the global minimum in the potential
energy. In contrast to this, the more rapid quench is highly skewed
with a second broad peak which can be observed at high values of $\Delta W_{\tau}/\beta$.
This is due to many of the trajectories becoming stuck in either the
local or global minima for prolonged times during the quench, \emph{i.e.}
loosely speaking, a break down in ergodicity. Due to the form of the
exponential average in Eq. \ref{JE} the long skewed wing and broad
second peak, for the more rapid quench, make only a small contribution
to the average. This is exactly compensated for by the trajectories
which remain trapped in the global minimum and have comparatively
little work done on them, but make a large contribution to the average.
Thus the Jarzynski Equality gives the same change in free energy that
would be obtained from a single trajectory that is quenched quasistatically.
Clearly these distributions are not Gaussian yet it is readily apparent
that the distribution will approach a Dirac delta function as the
quench time is extended towards infinity. 

\subsection{Quasiequilibrium free energy expressions}

\subsubsection{Quasiequilibrium free energy from the quasiequilibrium partition
function}

As seen in Fig. \ref{fig:distribution-functions} the distribution
function of the more rapidly quenched data is fitted very well at
the time of $\tau=200$, with $\epsilon=1$, by Eq. (\ref{dist_fn_quasiequilibrium})
with $a_{1}/a_{2}=0.03627$ ($a_{1}=0.07000;\: a_{2}=1.930$). Using
these values of $a_{\alpha}$ and numerical integration to evaluate
the local partition functions, ($Z_{\alpha}=\int d\boldsymbol{\Gamma}s_{\alpha}(\mathbf{\Gamma})\exp[-\beta H_{e}(\boldsymbol{\Gamma})])$,
Eq. (\ref{quasiHelmholtz_glass}) can be used to find that $\Delta A_{Z}=4.276$
when $\epsilon$ is changed from $\epsilon=0.1$ to $\epsilon=1$
over this period. We can then compare this result with the value obtained
using averages over nonequilibrium paths, Eq. (\ref{Nonequilibrium AZ expression})
where $\Delta X_{Z,\tau}$ is given by Eq. (\ref{work_glass}). This
method gives $\Delta A_{Z}=4.281$, and the values are obviously in
good agreement. It is worth noting that if we waited long enough for
the system to equilibrate and then calculated $\Delta X_{Z,\tau}$,
at this time $\tau$ (which is much longer than 200), we would once
again obtain the standard change in free energy, $\Delta A_{Z}=\Delta A_{eq}$.
The large difference between $\Delta A_{eq}$ (1.7219) and $\Delta A_{Z}$
(4.276) we find here shows how far the system is out of equilibrium
at time $\tau=200$ by this measure. 

We can now use $\Delta A_{Z}$ and Eq. (\ref{quasi-Helmholtz 1})
to determine the difference in Helmholtz free energy of the equilibrium
and quasiequilibrium states. With $a_{1}=0.07000,\: a_{2}=1.930$
and using numerical integration we find $\Delta A=1.960$, which is
significantly different from $\Delta A_{eq}$. This shows the importance
of using the new expressions for the quasiequilibrium free energy
if it is necessary to find the free energy of the glass state. In
cases where $a_{1}Z_{1}/a_{2}Z_{2}\backsimeq0$, it would be possible
to obtain a good approximation to this free energy by approximating
the potential energy of the glass state with Eq. (\ref{potential})
and $\epsilon=1$ for $q\ge0.14370505$, but with $\Phi=\infty$ for
$q<0.14370505$. However, for the data presented here $a_{1}Z_{1}/a_{2}Z_{2}\backsimeq8.657$
and this is not appropriate.

\begin{figure}
\resizebox{8.5cm}{!}{\includegraphics{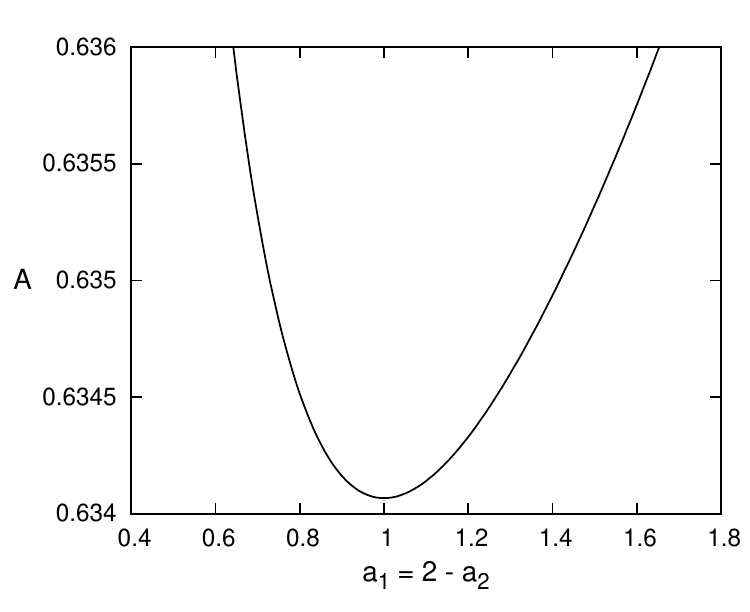}}

\caption{\label{fig:Free-energy}The free energy for quasiequilibrium systems
with a potential Eq. (\ref{potential}) with $\epsilon=$1 and the
distribution function given by Eq. (\ref{dist_fn_quasiequilibrium}).
A minimum can be seen at equilibrium, $a_{1}=a_{2}=1$. }

\end{figure}

In order to demonstrate that the Helmholtz free energy, $A$, given
by Eq. (\ref{Helmholtz in terms of AZ}) is minimised when $a_{1}=a_{2}=1$,
as discussed in Section III C, we plot $A$ as a function of $a_{1}=2-a_{2}$
for the potential, Eq. (\ref{potential}), with $\epsilon=1$ in Fig.
\ref{fig:Free-energy}. The numerical data clearly support the theoretical
result.

\subsubsection{Quasiequilibrium free energy from a weighted sum of local domain
free energies}

Above we have shown that the free energy can be computed using Eqs.
(\ref{Nonequilibrium AZ expression}) \& (\ref{Helmholtz in terms of AZ}).
However the approach has a serious drawback. If we wish to compute
the free energy of a realistic model glass former using molecular
dynamics simulations on a many body system, there will be a huge number
of ergodic domains that must be considered. The number of domains
will simply be too large to handle by these relations since it is
necessary to identify what domain each trajectory belongs to, and
to determine $a$ and various averages involving $a$ for each domain.
A similar problem will occur, in general, with use of Eqs. (\ref{conditional quasi <exp-beta.W>})
\& (\ref{NEFERQEqm}).

Here we devise an algorithm based on Eqs. (\ref{conditional-backward-glass-JE})
\& (\ref{backward-glass-JE}) that avoids this. We rewrite Eq. (\ref{backward-glass-JE})
as the following average. 

\begin{equation}
\Delta A^{qe\rightarrow eq}=-k_{B}T\overline{\ln\left[\left\langle \exp(-\beta\Delta W_{\tau}^{qe\rightarrow eq}(\mathbf{\Gamma)})\right\rangle _{\alpha_{qe}}\right]}\label{av backward-glass-JE}\end{equation}

The over bar means that we sample master points in the phase space
from the relaxed $f^{(2)}(\mathbf{\Gamma})$ distribution formed from
an ensemble of quenched systems. These master points will belong to
the various domains and will by definition populate those domains
according to the weights appearing in Eq. (\ref{backward-glass-JE})
namely $w_{\alpha}^{(2)}$. For each master point we calculate the
average $\left\langle \exp(-\beta\Delta W_{\tau}^{qe\rightarrow eq}(\mathbf{\Gamma)})\right\rangle _{\alpha_{qe}}$
over the domain that each master point resides in. This is done by
generating daughter points from their master. These daughter points
are guaranteed to belong to the same domain as their master, because
they are generated either by fixing the configuration from the master
point and sampling the momenta from the appropriate Maxwell-Boltzmann
distribution or by simply using the equations of motion to ergodically
generate points in that same domain. In this Monte-Carlo like procedure
for correctly averaging within and between domains we never need to
know how many domains there are, or what their explicit weights are.
The weights occur naturally and we know how to perform averages over
each ergodic subdomain.

In the numerical work considered above, a quench time of $\tau=200$
is sufficiently long that the numerical distribution is well approximated
by the quasiequilibrium distribution when the quench is complete,
and no additional relaxation time is required. This is demonstrated
by the fact that the quasiequilibrium distribution function gave an
excellent fit to the numerical data. Therefore Eq. (\ref{av backward-glass-JE})
could be applied using states generated at the end of the quench.
The free energy difference between the quasiequilibrium and equilibrium
states was determined using $N_{q}=5\times10^{4}$ trajectories or
master points generated using the $\tau=200$ quench. From each of
these, an ensemble of $5000$ trajectories was spawned to run in reverse
by sampling the appropriate Gaussian distributions for the momentum
$p$ and the thermostat multipliers $\zeta_{1}$ and $\zeta_{3}$
given by Eq. (\ref{distribution function}). Eq. (\ref{av backward-glass-JE})
was then used to calculate a change in free energy of $\Delta A=-\Delta A^{qe\rightarrow eq}=1.964\pm0.006$.
This value compares favourably with the above-mentioned directly obtained
value of $\Delta A=1.960$, demonstrating the validity of the technique.

\section{Conclusions}

In this paper we have considered a number of different states: 

\textbullet{} time dependent nonequilibrium states;

\textbullet{} ergodic equilibrium states where there is only one phase
space domain, say domain 1, for this domain , $a_{1}=1$ and within
this single domain phases are Boltzmann distributed;

\textbullet{} nonergodic quasiequilibrium states where the domain
weights $a_{\alpha}$ are time independent but essentially arbitrary
and the intra-domain weights are Boltzmann distributed; and lastly

\textbullet{} constrained equilibrium states where $a_{\alpha}=1,0\;\forall\;\alpha$
, and for the occupied domains both the intra and inter domain weights
are Boltzmann distributed.

By studying a simple model we have shown that subject to the ergodic
consistency condition, that by performing nonequilibrium path integrals,
the Jarzynski Equality can be used to predict free energy differences
between states that are either in thermodynamic equilibrium or constrained
thermodynamic equilibrium. We have shown that subject to this condition
this equality can even be used in systems where after the change protocol
between the two states has completed, the final relaxation to the
new equilibrium state is exceedingly slow. Our example also confirms
the correctness of a new statistical mechanical treatment\citep{Williams-JCP-07}
of time independent, nonergodic, nondissipative nonequilibrium systems
- so-called quasiequilibrium systems. 

We have also shown that in systems where certain phase space domains
are totally unsampled in the final observed distribution of states
($a_{\gamma}=0$), the Jarzynski Equality gives the free energy difference
between the equilibrium states and the final constrained equilibrium
state.

We have derived three variations of the Jarzynski Equality (Eqs. (\ref{quasi-Helmholtz 1}),
(\ref{NEFERQEqm}) \& (\ref{backward-glass-JE})) which calculate
free energy differences between the initial equilibrium state and
the final quasiequilibrium state. Due to the intractably large number
of domains to be considered, the first two of these new expressions
would be very difficult to use on a more realistic examples of a glass.
The third result can be restated as Eq. (\ref{av backward-glass-JE}),
which allows these problems to be overcome by providing the free energy
of the quasiequilibrium state directly in terms of averages, without
it being necessary to explicitly enumerate and characterise the domains.
Combined, these results provide a concise illustration of how thermodynamics
relates to glasses, polymorphs or similar arrested systems.

The approach we have described in the main text above considers ensembles
of quenched states, which in general will produce a number of different
ergodic domains, (e.g. glasses with different physical properties
or different polymorphs of a material). The free energies calculated
therefore generally refer to the free energy of this \emph{ensemble}.
From a practical perspective, it is often of more interest to obtain
the free energy of a single ergodic state (e.g. a single glass sample).
The free energy of this system would be equal to that of a quasiequilibrium
state where only one of the domains is populated. We note that Eqs.
(\ref{quasi-Helmholtz 1}) \& (\ref{NEFERQEqm}) could not strictly
be applied under these conditions, since ergodic consistency would
be violated - points in the initial equilibrium state might lead to
points in the final state that are not within the required domain.
However, Eqs (\ref{backward-glass-JE}) \& (\ref{av backward-glass-JE})
meet the required conditions, and $N_{D}=1,\; w_{1}^{(2)}=1$. Eq.
(\ref{av backward-glass-JE}) takes on the particularly simple form
$\Delta A^{\alpha\rightarrow eq}=-k_{B}T\ln\left[\left\langle \exp(-\beta\Delta W_{\tau}^{\alpha\rightarrow eq}(\mathbf{\Gamma)})\right\rangle _{\alpha}\right]$
where \textit{$\alpha$} refers to the phase space domain that characterises
the sample. This approach allows the free energy of individual ergodic
subdomains to be determined relative to an ergodic equilibrium state.

\begin{acknowledgments}
DJE would like to thank the Isaac Newton Institute for Mathematical
Sciences, Cambridge UK, for support and Chris Jarzynski for useful
early discussions on this topic at that Institute. DJS would like
to thank Chris Jarzynski and his group for hosting her visit to the
University of Maryland and discussion of this work. SRW, DJS and DJE
would like to thank the Australian Research Council for support for
this research. 
\end{acknowledgments}

\begin{thebibliography}{25}
\bibitem[{end()}]{endnote-1}
\bibinfo{note}{Note the path is thermostatted but the derivative is computed as
  the adiabatic derivative.}
\end{thebibliography}

\end{document}